# Title: Intrinsic topological spin probes for electrical imaging of nanoscale energy landscapes

**Authors:** Liam K. Mitchell[1], Benjamin J. Brown[1], Gang Xiao[1]*

**Affiliations:**

[1]Department of Physics, Brown University; Providence, RI, 02912, USA.

*Corresponding author. Email: gang_xiao@brown.edu

**Abstract:** Disorder in magnetic materials prevents reliable control of spin textures and constrains their integration into spintronic devices. Existing methods access disorder only indirectly through external imaging probes or bulk transport measurements, leaving the internal energy landscape inaccessible. We introduce an intrinsic magnetic microscopy method in which a topological spin texture serves as a mobile probe of disorder, directly mapping energy landscapes inside multilayer devices without probe-sample separation. Using a ~10-nm magnetic vortex core confined within a magnetic tunnel junction, we track its displacement with nanometer-scale sensitivity to resolve intrinsic and engineered defect-induced potentials and directly quantify local pinning forces. This framework establishes spin textures as internal spectroscopic probes of disorder and enables quantitative engineering of pinning structures in functional magnetic systems.

**Main Text:** Nanoscale spin textures such as domain walls, skyrmions, and magnetic vortices offer a route toward unconventional computing and energy-efficient spintronic technologies (*1–4*). Although these textures can be reliably nucleated and driven by magnetic fields or electrical currents, their motion is governed by nanoscale disorder that locally modifies the magnetic free energy and gives rise to pinning, trajectory deflections, and stochastic motion (*5–7*). As device dimensions approach intrinsic magnetic length scales (*8*), uncontrolled disorder increasingly limits deterministic operation and reproducibility in practical spintronic architectures.

However, the microscopic energy landscape governing spin texture motion inside operational devices remains experimentally inaccessible. Existing approaches either visualize spin configurations near defects (*9–15*) or infer pinning indirectly through macroscopic transport signatures including critical current thresholds, resonance shifts, and hysteresis (*16–19*). These measurements capture global responses rather than the local forces acting on the spin texture itself, leaving pinning strength and spatial inhomogeneity unresolved. This limitation is especially acute in multilayer stacks where buried layers are isolated from surface-sensitive microscopies (*20*), creating a fundamental obstacle to both deterministic control and harnessing disorder as a functional resource (*21*).

Here we introduce a magnetic metrology strategy in which a confined topological spin texture serves as an internal probe of the energy landscape within an operational device. A nanometer-scale vortex core confined in a magnetic tunnel junction (MTJ) can be displaced deterministically by applied magnetic fields and tracked electrically through spin-dependent tunneling. Because the probe resides within the MTJ and samples the local potential within the magnetic film (*22*), its motion directly encodes the underlying disorder without probe-sample separation. Unlike prior studies that infer vortex pinning from gyrotropic resonance shifts (*23*,

*24*) or localized hysteresis (*25*), our approach reconstructs the effective pinning potential from controlled core displacement, enabling quantitative mapping of local defect forces over large two-dimensional areas. We show that quantized core dynamics reveal a richly structured and reproducible defect map within nominally uniform films, with pinning forces evolving systematically from 10 to 300 K and engineered defect landscapes that map one-to-one onto device topography. These results transform confined topological spin textures from passive dynamical elements into active probes of the energy landscapes governing collective dynamics.

**Device architecture and electrical vortex readout**

To realize an internal, electrically accessible probe of disorder, we employ an MTJ whose free layer stabilizes in a single magnetic vortex (*26*, *27*). In this architecture, the vortex core (*28*) acts as a confined, particle-like object whose position can be deterministically controlled by an in-plane field $H_x$, while spin-dependent tunneling provides a direct electrical readout of its motion. Because the core resides within the active magnetic layer, its displacement encodes the local magnetic free energy inside the device, eliminating the need for external probes or mechanical motion.

The MTJ stack consists of $Co_{40}Fe_{40}B_{20}$(3.5 nm)/MgO(2.9 nm)/$Co_{40}Fe_{40}B_{20}$(55 nm) multilayers grown on oxidized silicon and patterned into circular junctions with diameters $d$ = 10-20 μm (Fig. 1A). The 55-nm thick free layer stabilizes a single vortex state (*29*), while post-growth magnetic annealing produces high tunnel magnetoresistance (TMR) ratios of 110% at 300 K. Magnetoconductance $G$ is measured under a 100-mV dc bias as a function of $H_x$, which is linearly mapped to the magnetization $M(H_x)$ (see Materials and Methods).

Across 10-300 K, the magnetization curves exhibit linear, non-hysteretic behavior characteristic of a single vortex core (*30–32*) (Supplementary Text). Near zero field, the vortex core translates perpendicular to $H_x$ (Fig. 1B), with positional susceptibility (*33*):

$$\chi = \frac{dY}{dH_x} \approx C \frac{d^2}{M_{sat} t}$$

where $M_{sat}$ is the free-layer saturation magnetization and $t$ is its thickness. Weak logarithmic corrections associated with disk aspect ratios are absorbed into the prefactor, from which we experimentally determine $C$ = 0.36. For a device with $d$ = 15 μm, this yields $\chi$ = 98.7 nm/Oe, establishing nanometer-scale positioning of the core. The reversible operating range is bounded by the vortex annihilation field $H_{an}$, which scales inversely with disk diameter $d$ (*34*). For our MTJs, we obtain $H_{an}$ = 1125 μm · Oe/$d$ (Supplementary Text). This scaling defines the field window over which the vortex core can be positioned without irreversible switching, thereby setting the dynamic range for disorder mapping.

**Discrete vortex motion in pinning landscapes**

We first characterize the quasistatic motion of the vortex core to reveal its interaction with intrinsic disorder in the CoFeB film. In a defect-free disk, the vortex core is expected to translate smoothly and reversibly under $H_x$. Instead, the measured $M(H_x)$ curves exhibit a linear background punctuated by discrete steps with amplitudes of $10^{-4}$–$10^{-3}\Delta M/M_{sat}$ (Fig. 1C). These discontinuities sharpen significantly upon cooling from 300 K to 10 K, while their field positions remain reproducible across successive sweeps (Supplementary Text), indicating that they originate from interactions with a fixed pinning landscape rather than noise or thermal fluctuations.

These step features demonstrate that the spin texture encounters variations in the magnetic free energy as it translates through the sputtered film (Fig. 1D). To identify their microscopic origin, we performed micromagnetic simulations (*35*) incorporating a randomized distribution of defects modeled as reductions in magnetic parameters (see Materials and Methods). We first verified that defects positioned several core radii away from the core's trajectory exhibit negligible influence on the global magnetization (Supplementary Text). After establishing this locality, defects were placed along the core path, where the hysteresis loops reproduce the experimentally observed steps (Fig. 1C).

Visualization of the simulated core trajectory reveals that each large magnetization step corresponds to an abrupt hopping of the core between neighboring pinning sites (Fig. 1E, Movie S1). Between these transitions, the core remains trapped while the surrounding vortex structure deforms elastically under $H_x$. These two dynamical regimes, abrupt depinning and elastic pinning, constitute the fundamental signatures of defect-mediated vortex motion and are analogous to behaviors observed in other spin textures (*36*, *37*). The measurement and simulations demonstrate that the electrical signal encodes the sequence of metastable states traversed by the core, providing real-space access to the magnetic energy landscape.

**Statistical signatures of vortex pinning**

To quantify the depinning and pinning modes of the vortex, we performed a statistical analysis of the magnetization increments $\Delta M/M_{sat}$ under a field step of $\Delta H_x = 80$ mOe. For the $d = 15$ μm device, this corresponds to an average core displacement of $\Delta Y = \chi \Delta H_x = 8$ nm, providing spatial resolution comparable to the characteristic size of the individual pinning sites. Histograms of magnetization step amplitudes exhibit a pronounced bimodal distribution (Fig. 2B), well described by the sum of two log-normal components. This structure indicates the coexistence of two distinct modes of vortex motion.

The narrow, small-amplitude population corresponds to elastic pinned motion whereas the larger amplitude population reflects core-depinning events between metastable sites (Fig. 2A). To estimate their characteristic energy scales, we convert the magnetization changes into effective Zeeman energy variations. At 300 K, pinning and depinning events correspond to energy changes of $E_p = 30.2 \pm 0.2$ meV and $E_d = 74.1 \pm 4.3$ meV, respectively, comparable to the thermal energy $k_B T = 25$ meV. In this regime, thermal fluctuations significantly assist barrier crossings, leading to comparable populations of each mode (Fig. 2C).

At 10 K, the depinning energy increases to $E_d = 98.8 \pm 2.4$ meV, while $k_B T$ decreases to 0.9 meV. The strong separation between barrier energy and thermal energy suppresses thermally assisted hopping, consistent with Arrhenius-type activation over defect-induced barriers (*25*). As a result, depinning becomes predominantly field driven, producing fewer but larger core displacements. Pinned motion therefore dominates at 10 K, accounting for ~65% of all events (Fig. 2C). The pinned energy scale ($E_p \sim 30$ meV) remains temperature independent, consistent with elastic deformation of the vortex structure weakly influenced by thermal fluctuations.

To confirm whether these statistics depend on specific material details, we performed ensemble simulations incorporating randomized defects which reproduce the experimentally observed bimodal log-normal structure (Fig. 2D). These results establish pinning and depinning as universal modes of vortex dynamics in pinning landscapes that can be resolved quantitatively through purely electrical measurements.

**Microscopic origin of vortex-core pinning**

To identify the microscopic origin of vortex-core pinning, we examine how a localized spin texture interacts with variations in magnetic material properties. Because the magnetization gradients of the vortex are largest within the core, its energetics are strongly influenced by the exchange energy density (Supplementary Text):

$$w_{ex} = A_{ex}|\nabla \vec{m}|^2$$

where $A_{ex}$ is the exchange stiffness and $\vec{m}$ is the unit magnetization. A local reduction in $A_{ex}$, or in other parameters that influence spin canting, lowers the energetic cost of strong gradients and creates an attractive potential minimum for the core. Additionally, a reduction in $M_{sat}$ decreases the demagnetization energy associated with an out-of-plane core in a thin disk, further stabilizing it at a defect site. Because both the exchange energy and demagnetization energy are sharply localized around the core—decaying over a length scale comparable to the exchange length—defects located away from the core have negligible effects on the total vortex energy.

For representative defects in our devices, the energy reduction associated with defect-core overlap is on the order of ~200 meV (next section), whereas the Zeeman energy corresponding to a field increment of $\Delta H_x$ = 80 mOe is approximately ~50 meV. When the defect-induced energy reduction exceeds the incremental Zeeman drive, the vortex core becomes trapped rather than translating smoothly. This controlled energy separation enables defect resolution and motivates our chosen step size; larger field increments would allow the core to bypass individual barriers.

In sputtered films, such magnetic reductions can arise from grain boundaries, thickness variations, atom-scale dislocations, and nonmagnetic inclusions (*38–40*). In the limiting case of a material void used in simulations ($A_{ex}$ and $M_{sat}$ set to zero), the strong energy reduction produces robust pinning sites that generate magnetization steps consistent with experiment.

**Reconstruction of defect-induced pinning potentials**

Moving beyond the identification of vortex dynamics, we resolve the underlying pinning landscape experienced by the core by reconstructing the pinning potential in a real device. Simulations first establish the expected interaction between the core and individual defects. By sweeping the core across isolated nanoscale defects and calculating the exchange energy, we obtain attractive wells with characteristic depths of 120 meV and lateral extents comparable to the core radius (Fig. 3A and B). The minimum of the wells coincides with the defect positions, confirming that the sensing region of the vortex is spatially defined by the ~10-nm core.

Experimentally reconstructing the interaction potential requires separating defect-induced energy variations from the smooth magnetostatic confinement of the vortex state (Supplementary Text). The total magnetic free energy $E(Y)$ is obtained by integrating the magnetization response and using the small field relation between $H_x$ and core displacement $Y$ (see Eq. 1). To isolate disorder contributions, we subtract the reference energy of an ideal vortex described by the rigid vortex model (*41*). In the small-field reversible regime, deformable models (*42*) introduce only a renormalization of the harmonic stiffness, defining a smooth background energy:

$$E(Y) = \frac{1}{2}kY^2 - h\left(Y - \frac{Y^3}{8}\right)$$

where the first term represents the magnetostatic restoring force and the second term captures the non-linear Zeeman coupling. Deviations from this reference encode the magnetic energy variations experienced by the vortex core due to local inhomogeneities, thereby isolating the residual pinning potential $U_{\text{pin}}(Y)$.

Applying this reconstruction to simulations reveals that interactions between nearby defects form composite pinning structures consisting of overlapping wells with enhanced depths (Fig. 3C). This reveals that nanoscale defects can aggregate into effective pinning clusters that extend over tens to hundreds of nanometers with wells depths approaching 400 meV, despite individual defects remaining nanoscale. The magnetization curve and its derivative show that minima in $U_{\text{pin}}(Y)$ align with regions of pinned motion, linking the reconstructed wells directly to vortex-core dynamics.

Applying the procedure to experimental data yields pinning potentials with closely analogous features (Fig. 3D). The reconstructed landscapes consist of broad wells with depths of approximately 150-300 meV, indicating strong confinement of the vortex core by intrinsic disorder in the CoFeB film. The spatial positions of these wells remain fixed between 10 and 300 K, demonstrating a device-specific disorder landscape rather than thermally fluctuating features.

While the well locations are temperature independent, their morphology evolves systematically. At low temperatures the wells are deeper and more sharply curved, consistent with reduced thermal activation. Quantitatively, the local curvature (pinning stiffness) of the dominant well at $Y = 1.5$ μm follows an exponential temperature dependence, $\alpha \propto \exp(-T/T_0)$, with $T_0 = 108$ K (inset of Fig. 3D). These results confirm that the vortex core functions as a probe capable of reconstructing nanoscale pinning potentials in an operational device.

**Two-dimensional electrical mapping of defect landscapes**

Having established the vortex core as a localized probe of defect-induced pinning, we extend this approach to spatially resolve the full two-dimensional disorder landscape within the buried free layer. Independent control of orthogonal in-plane magnetic fields, $H_x$ and $H_y$, enables deterministic rastering of the vortex core across the disk without mechanical motion (Fig. 1B). As the core is translated, the magnitude of discrete magnetization jumps provides a local measure of pinning strength and energy gradients. Large discontinuities correspond to abrupt depinning transitions between adjacent potential minima, whereas smaller steps reflect elastic motion within a single well. By assigning these transport signatures to their corresponding spatial coordinates, the electrical signal is converted into a two-dimensional contrast map of the effective pinning potential.

Micromagnetic simulations validate the fidelity of this approach. For a simulated disk containing an array of prescribed defects, the resulting maps resolve each defect as a compact, high-contrast feature precisely localized at its true position. (Fig. 4A). Dark regions of reduced susceptibility indicate strong pinning at defect locations, whereas bright regions correspond to depinning into or away from the defect-induced potential well. Importantly, the lateral extent of each pinning event matches the vortex core radius, indicating that the spatial resolution is fundamentally limited by the core size. Outside the defect regions, the vortex maintains a constant, moderate susceptibility indicative of defect-free disks, confirming that only the core couples to local magnetic variations.

We apply this technique experimentally to a $d = 12$ μm MTJ. The resulting maps reveal a rugged and feature-rich disorder landscape composed of nanoscopic pinning sites and extended defect clusters (Fig. 4B and C). Sampling the device with a pixel spacing (set by the field step and device diameter, see Eq. 1) of approximately 25 nm, which is comparable to the intrinsic core radius, uncovers spatial features that are inaccessible to existing magnetic imaging techniques or external probes. Repeated scans yield reproducible maps, confirming that the observed features represent a static, device-specific disorder manifold (Supplementary Text).

To further validate this method, we introduce engineered pinning sites by etching two 25-nm deep pits into the $d = 12$ μm free layer using Ar ion milling (Fig. 4D). The reconstructed map (Fig. 4F and G) exhibits a one-to-one spatial correspondence with the artificial defect topography observed by scanning electron microscopy, accounting for the orthogonal relationship between applied field direction and vortex displacement. Using the known defect size and location, we confirm the core's displacement rate as $dY/dH_x = dX/dH_y = 62.5$ nm/Oe in both directions, which is used to determine the constant in Eq. 1. Line profiles through the landscape reveal two symmetric pinning wells of approximately 600 meV centered at the defect locations, substantially stronger than those associated with intrinsic disorder (Fig. 4E).

While the spatial resolution (~10 nm) is comparable to leading magnetic imaging techniques, its distinguishing capability lies in enabling purely electrical access to the internal energy landscape of buried, fully operational multilayer devices without probe-sample separation. Conventional magnetic microscopies primarily resolve surface spin configurations or stray-field distributions and often require optical access, vacuum environments, or mechanically scanned probes. Here, the vortex core operates entirely within the device architecture under standard electrical bias in regimes where external probes are impractical. The advance therefore lies in introducing a metrological framework in which a controllable topological excitation serves as a device-integrated probe of the local energy landscape governing collective dynamics.

**Discussion and outlook**

This work establishes an embedded approach for nanoscale magnetic metrology in which a topological spin texture serves as an all-electrical probe of spatial energy landscapes. By exploiting the sensitivity of a magnetic vortex core to local variations in magnetic material parameters, we reconstruct defect-induced pinning potentials and map intrinsic and engineered disorder within fully operational MTJs. The vortex core behaves as a controllable quasiparticle whose finite internal structure sets the spatial resolution; in principle, this resolution can be tuned by modifying geometry and magnetic length scales through material choice, thickness, or out-of-plane fields.

Beyond the vortex implementation, the central result is methodological: controlled motion of a localized excitation can be inverted to recover the energy landscape that governs its dynamics. In this sense, the measurement transforms transport observables into a spatially resolved map of pinning forces and metastable structure. This inversion framework should generalize to other controllable spin textures, provided they can be driven and detected with sufficient precision. More broadly, it introduces a form of disorder spectroscopy for spintronic stacks, enabling quantitative links between nanoscale inhomogeneity and device-scale behavior, including stochastic motion, noise, and switching margins.

An important implication is that the reconstructed landscapes are not merely diagnostic. Intrinsic landscapes encode device-specific microstructure, whereas engineered landscapes can be designed by introducing controlled heterogeneity. Without committing to specific device architectures, this capability motivates several directions: (i) disorder-based device fingerprinting for hardware identification; (ii) monitoring landscape evolution as a sensitive indicator of environmental exposure or cumulative damage; (iii) encoding information in engineered pinning architectures; and (iv) quantifying landscape drift under stress as a metric of aging and reliability. In each case, the enabling principle is the same: an internal probe that reconstructs an energy landscape electrically within an operating device.

Together, these results expand the role of topological excitations in spintronics from passive dynamical degrees of freedom to metrological elements that reveal and potentially enable control over the energetic structure governing collective motion. The ability to reconstruct these landscapes in situ establishes a foundation for quantitative studies of driven dynamics in complex potentials and for systematic engineering of disorder in nanoscale magnetic devices.

**Acknowledgments:**

AI-based tools were used solely for grammar and spell-checking during manuscript preparation. These tools did not contribute to the research design, interpretation, or generation of new ideas. All text modified with such assistance was independently reviewed and verified by the authors.

**Funding:**

This work was supported by (1) the Foundations of Superconducting Digital Logic (FSDL) program, Army Research Office/Devcom grant W911NF-24-1-0147 and (2) the National Institute of Biomedical Imaging and Bioengineering (NIBIB) of the National Institutes of Health (NIH) under Award No. UG3EB034695. The content presented here reflects the views of the authors and not necessarily those of the Army Research Office or NIH.

**Author contributions:**

Conceptualization: LKM, BJB, GX

Methodology: LKM, BJB, GX

Investigation: LKM, BJB, GX

Funding acquisition: GX

Supervision: GX

Writing – original draft: LKM

Writing – review & editing: BJB, GX

**Competing interests:** Authors declare that they have no competing interests.

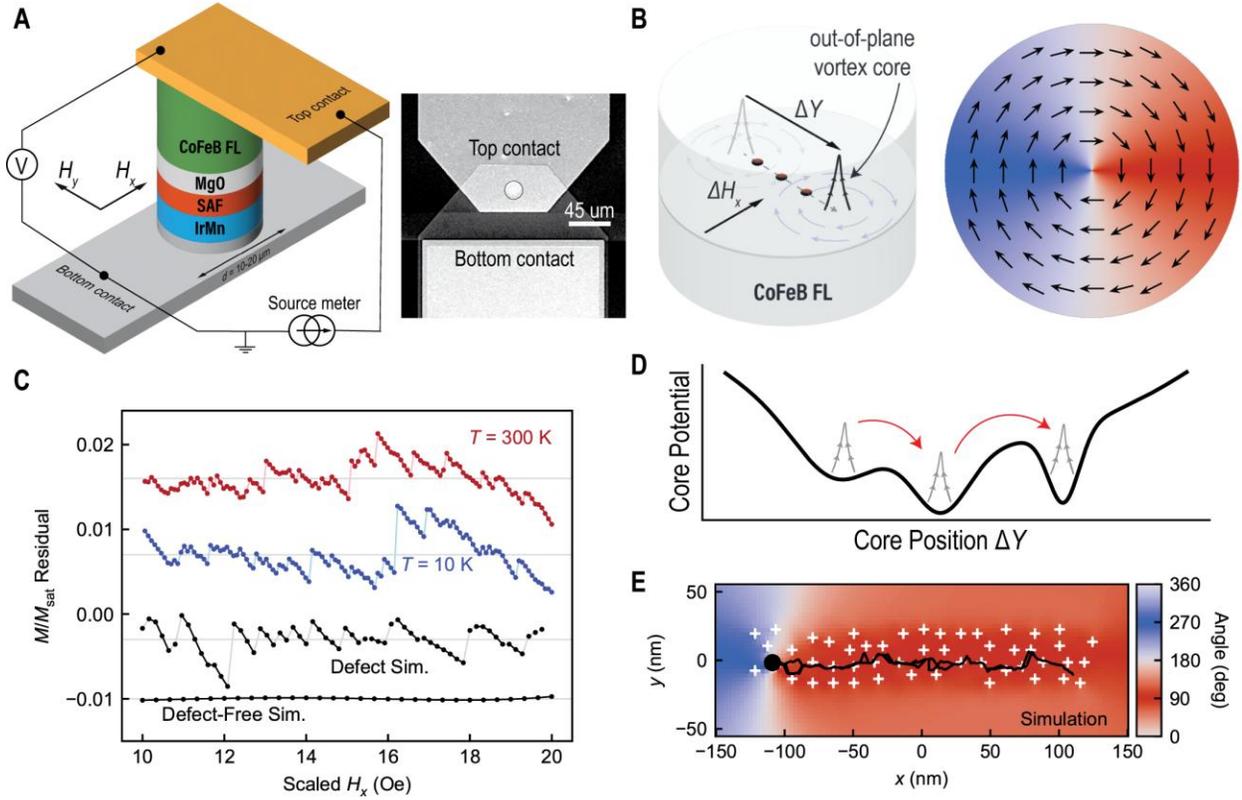

**Fig. 1. Electrical access to vortex-core dynamics enables probing nanoscale disorder.** (**A**) Left: Magnetic tunnel junction schematic and measurement configuration. A vortex state with a single form in the CoFeB free layer (FL), while the synthetic antiferromagnet (SAF) defines the magnetic sensing axis. Right: Scanning electron microscope image of a 15 µm-diameter MTJ. (**B**) Left: In-plane magnetic fields orthogonally displace the ~10-nm vortex core (out-of-plane $M_z$ region) throughout the device. Right: Top-down view of the zero-field vortex state (arrows indicate spin direction). (**C**) Background-subtracted magnetization traces at 10 K and 300 K reveal step-like discontinuities corresponding to core hopping events (shown in lighter color line segments). Micromagnetic simulations demonstrate that such jumps emerge only in the presence of defects, identifying disorder-induced pinning as their origin. Gray line denotes offset from zero for visual clarity. (**D**) Disorder-induced energy landscape seen by the vortex core. The core traverses the film by hopping between potential minima (red arrows). (**E**) Top-down view of simulated vortex-core trajectory (black line) through a random defect landscape (white crosses) extracted from a major loop. Hysteretic hopping between metastable minima produces quantized transport. The color map encodes the in-plane magnetization angle (see (**B**)); the larger black marker denotes the instantaneous core position.

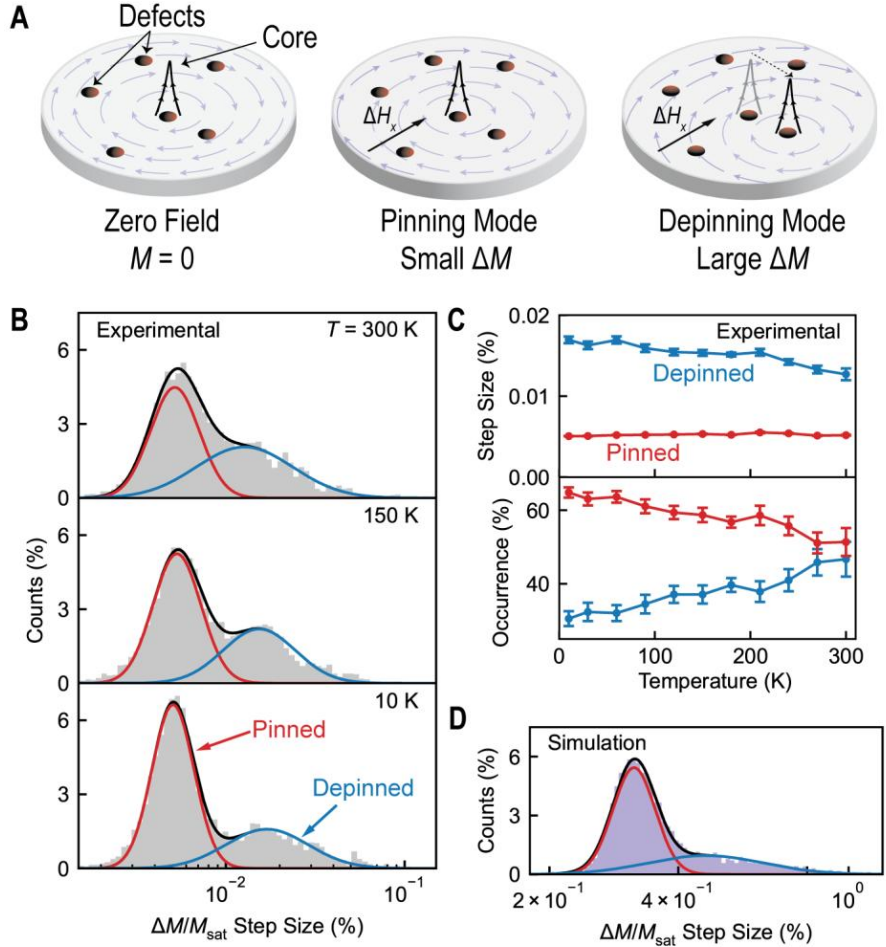

**Fig. 2. Bimodal statistics arising from discrete vortex-core transport in a disordered energy landscape.** (**A**) Schematic of defect-mediated vortex transport (defects shown as red circles). Left At zero applied field, the vortex core occupies a local energy minimum. Center: Under small field perturbations, the core undergoes elastic displacement within a single pinning well (pinned mode). Right: At larger drive, the core escapes the local minimum and hops to a neighboring metastable state (depinning mode), producing a larger magnetization discontinuity. (**B**) Histogram of experimentally resolved magnetization step amplitudes in the CoFeB free layer. The distribution separates into two log-normal components corresponding to two dynamical channels: small-amplitude elastic motion within a pinning well (pinned mode, red) and larger barrier-crossing events (depinning mode, blue). (**C**) Temperature evolution of the two dynamical modes. Top: The characteristic scale of the pinned mode is temperature independent, consistent with local harmonic confinement. In contrast, the depinning mode shifts to larger amplitudes at lower temperature, reflecting reduced thermal activation and stabilization of deeper energetic minima. Bottom: Relative populations extracted from bimodal fits. At low temperature, motion is dominated by elastic confinement, whereas elevated temperature increases the probability of activated depinning. (**D**) Statistical analysis of simulated *M(H)* traces in a randomized defect landscape. The spontaneous emergence of the same bimodal distribution

demonstrates that the two transport channels are universal consequences of a topological vortex core navigating a rugged disorder potential.

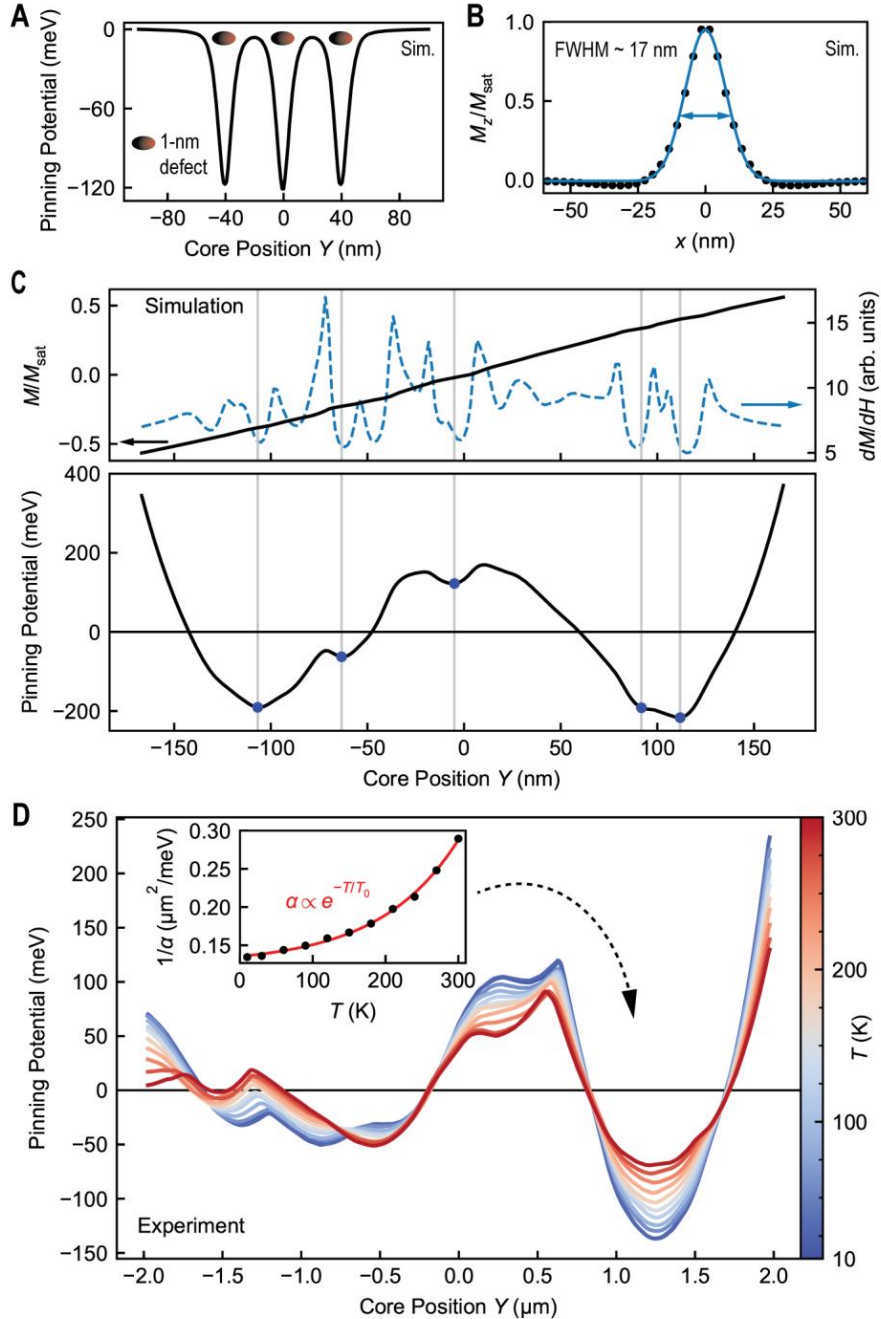

**Fig. 3. Direct reconstruction of the vortex-core pinning potentials.** (**A**) Calculated interaction potential for a vortex core coupled to three 1-nm defects. Each defect generates a narrow, attractive well (~120 meV), illustrating disorder-induced energy minima. (**B**) Micromagnetic simulation of the vortex core out-of-plane $M_z$ profile. The core is well described by a Gaussian with full width at half maximum of 17 nm, defining the intrinsic spatial resolution of the probe. (**C**) Top: Simulated $M/M_{sat}$ as a function of core position $Y$ in a randomized defect landscape (from Fig. 1D). The dashed blue curve shows the corresponding differential susceptibility $dM/dH$ (right axis, arbitrary units). Bottom: Reconstructed pinning potential $U_{pin}(Y)$, revealing a multiscale landscape of metastable minima. Each potential minimum corresponds with low

susceptibility regions (pinned motion) of the magnetization curve. (**D**) Experimentally reconstructed pinning potential extracted from measured $M(H_x)$ loops. Broad wells with depths of ~100-300 meV appear at fixed spatial positions with pinning strength evolving with temperature, identifying a static, device-specific disorder manifold. Inset: Temperature dependence of the curvature of the local well at $Y = 1.5$ μm, which increases exponentially with decreasing temperature.

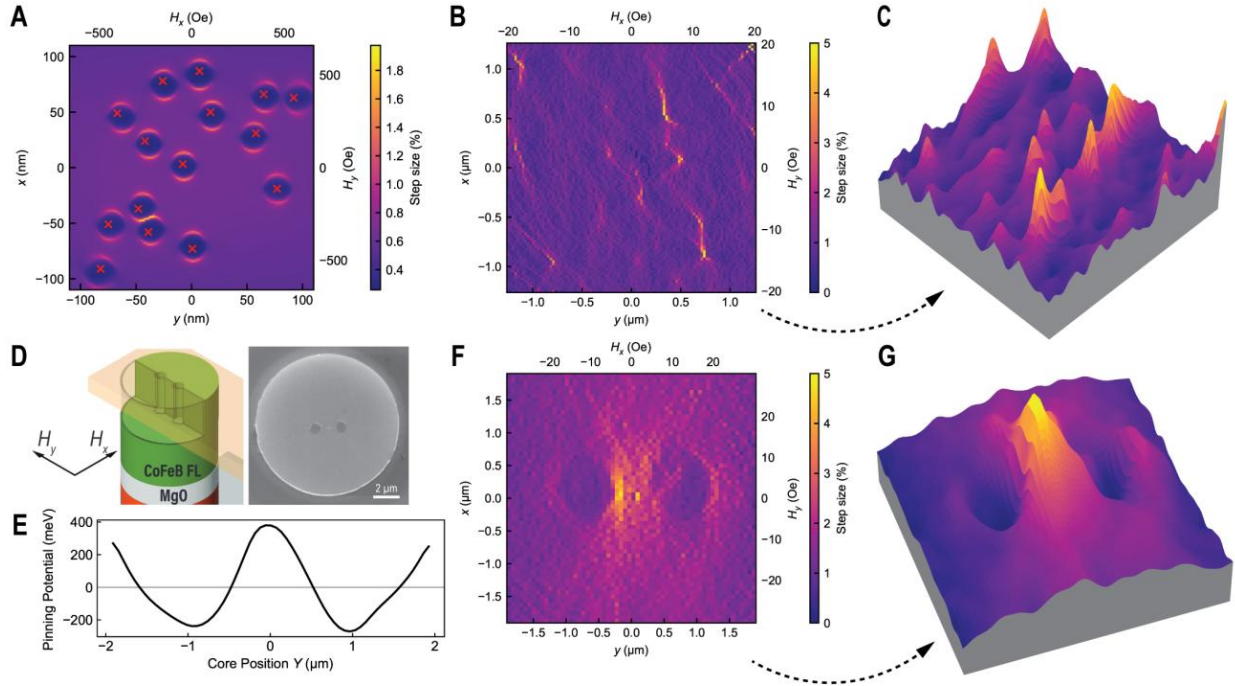

**Fig. 4. Electrical imaging of material disorder with an internal vortex-core nanoscale probe.** (**A**) Simulated two-dimensional defect map of a $d$ = 0.5 um disk containing engineered pinning sites (red crosses). Individual defects are resolved with spatial extent comparable to the intrinsic vortex-core diameter, establishing nanoscale imaging resolution. Color scale is differential magnetization along $H_y$. (**B**) Experimental two-dimensional map of a $d$ = 12 um MTJ. The reconstructed landscape reveals complex intrinsic disorder, including localized hotspots and extended defect clusters. Color scale represents the differential magnetoconductance along $H_x$. (**C**) Three-dimensional surface representation of the map shown in (**B**). (**D**) Device schematic (left) and scanning electron micrograph (right) of an MTJ containing two 25-nm deep artificial pinning pits defined by Ar ion milling. (**E**) Reconstructed pinning potential of the artificial defects along $H_y$ = 0 Oe. (**F**) Electrical 2D scan of the device in (**D**). The defect map exhibits one-to-one spatial correspondence with engineered features visible in the SEM image, resolving both engineered features and background disorder. (**G**) Three-dimensional surface representation of the map shown in (**F**).